\documentclass[aps,floatfix,nofootinbib,prd, notitlepage,11pt]{revtex4-1}

\usepackage{graphicx}
\usepackage{epic}
\usepackage{eepic}
\usepackage{latexsym}
\usepackage{amssymb,amsmath}
\usepackage{upgreek}

\newcommand{\eq}[1]{(\ref{#1})}
\newcommand{\be}{\begin{equation}}
\newcommand{\ee}{\end{equation}}
\newcommand{\bea}{\begin{eqnarray}}
\newcommand{\eea}{\end{eqnarray}}

\newcommand{\hs}[1]{\hspace{#1 mm}}

\newcommand{\ls}{{\bf L}_s}
\newcommand{\zs}{\zeta_{sh}}

\newcommand{\mt}{\tilde{m}}

\def\d{\delta}

\def\e{\epsilon}

\def\f{\phi}
\def\fr{\frac}

\def\l{\lambda}

\def\m{\mu}
\def\n{\nu}

\def\z{\zeta}
\def\O{\Omega}

\def\y{\eta}
\def\o{\omega}

\def\del{\partial}

\let\bm=\bibitem
\def\nn{\nonumber}

\begin{document}


\title{Adiabatic regularization of functional determinants in cosmology and radiative corrections during inflation} 

\author{Ali Kaya}

\email[]{ali.kaya@boun.edu.tr}

\author{Emine Seyma Kutluk}

\email[]{seymakutluk@gmail.com} 

\affiliation{Bo\~{g}azi\c{c}i University, Department of Physics, 34342, Bebek, \.{I}stanbul, Turkey}

\date{\today}

\begin{abstract}

We express the in-in functional determinant giving the one-loop effective potential for a scalar field propagating in a cosmological spacetime in terms of the mode functions specifying the vacuum of the theory and then apply adiabatic regularization to make this bare potential finite. In this setup, the adiabatic regularization offers a particular renormalization prescription that isolates the effects of the cosmic expansion. We apply our findings to determine the radiative corrections to the classical inflaton potentials in scalar field inflationary models and also we derive an effective potential for the superhorizon curvature perturbation $\zeta$ encoding its  scatterings with the subhorizon modes. Although the resulting modifications to the cosmological observables like nongaussianity turn out to be small, they distinctively appear after horizon crossing.  

\end{abstract}

\maketitle

\section{\leftline{Introduction}}

Functional determinants arise in various instances in quantum field theory like in the calculations of the effective actions, gauge fixing Faddeev-Popov terms, semiclassical tunneling amplitudes and Jacobian factors (for a review see e.g. \cite{r1}). In general, they appear as results of Gaussian path integrals and there are different methods in evaluating them such as the heat-kernel expansion, zeta function regularization and  Gel'fand-Yaglom theorem.  

The functional determinants also appear in quantum cosmology since Gaussian in-in/Schwinger-Keldysh path integrals are often encountered. For example, in certain models of preheating one assumes the existence of a reheating scalar that interacts with the inflaton, yet the action may still be quadratic in the reheating scalar yielding a Gaussian path integral in quantum theory (see \cite{reh1,reh2}). Similarly, in single scalar field models one may consider integrating out the quadratic inflaton fluctuations about the inflating background to determine the quantum  backreaction effects (see e.g.\cite{pi1,pi2,pi3,pi4,pi5,pi6}). Obviously, the functional determinants in a cosmological setting are time dependent and they can naturally be interpreted as effective actions.  

Most of the time the physically relevant quantity is not the sole functional determinant of an operator but the ratio of the  determinants of two closely related operators. In \cite{kc} a simple formula that gives the value of a normalized Gaussian path integral with a time dependent frequency is derived and we adapt that formula to the cosmological in-in functional determinants in the appendix (a formally similar trick has been frequently used in cosmological applications but the derivation of \cite{kc} is rigorous). It turns out that in a cosmological setting the result can be expressed in terms of the mode functions identifying the vacuum of the theory. This simple observation allows one to work out the Gaussian path integrals in several cosmological backgrounds with different vacua and utilize various approximation methods. We show that the bare Coleman-Weinberg effective potential \cite{cw} can be obtained by using the standard mode functions of the flat space. 

The effective potential obtained after a Gaussian integral involves a single loop momentum integral and it can be identified as a one-loop approximate result that is otherwise exact. Typically, this formal expression has the standard ultraviolet (UV) divergence which must be cured by a viable method. On the other hand, in (quasi) de Sitter space one also encounters peculiar infrared (IR) divergences in loop calculations (see e.g. \cite{ir1,ir2}). As discussed in \cite{ir3}, different regularizations agree in de Sitter space when the UV and IR cutoffs are chosen to be constants in physical and comoving spaces, respectively.

The dimensional regularization has  been used extensively to calculate the one-loop effective potentials in de Sitter space, see e.g. \cite{pi1,pi2,pi3,pi4,pi5,pi6} (see also \cite{y1,y2,y3,y4} for different approaches).  Therefore, in this paper we prefer to utilize adiabatic regularization \cite{ad1,ad2} in curing the loop infinities of cosmological functional determinants. Recently, the adiabatic regularization has been used to renormalize the inflationary power spectrum revising some of the well known results \cite{park1,park2,park3}. We believe that in the absence of any direct observational data, it is useful to compare different regularization methods to have a clearer picture of the physics. 

As we will discuss below, the adiabatic regularization captures the contribution of the cosmic expansion on the quantum effective potential. Namely, it encodes ``particle creation effects" rather than measuring ``vacuum polarization". We show that the terms subtracted by the adiabatic regularization can be interpreted as counterterms appearing in the bare action. Moreover, the calculation is IR safe, i.e. no IR divergences appear that require a careful treatment.    

In this paper, we also consider the coupling of the superhorizon curvature perturbation $\zs$ to the subhorizon modes of a scalar field in a quasi-de Sitter spacetime with constant deceleration. Since $\z$ is a special metric function, the coupling we are dealing with is fairly universal. By integrating out the Gaussian subhorizon modes in the path integral, it is possible to obtain an effective potential for $\zs$ encoding its scatterings with subhorizon modes. We determine how this quantum effective potential affects the power spectrum and nongaussianity. As expected, the corrections turn out be suppressed by the factor $H^2/M_p^2$ but they also appear after horizon crossing, which is a distinguishing feature. 

\section{\leftline{A Sample Problem}}

Consider two real scalar fields $\f$ and $\chi$ that are minimally coupled to gravity with the standard action
\be\label{a}
S=-\fr12\int d^4x\sqrt{-g}\left[\nabla_\m\phi \nabla^\m\phi+\nabla_\m\chi \nabla^\m\chi+2 V(\phi,\chi)\right],
\ee
where the potential is assumed to have the following form 
\be\label{pot} 
V(\f,\chi)=v(\f)+\fr12 \tilde{m}^2\chi^2+\fr12 g^2\phi^2\chi^2. 
\ee
The scalars are taken to be propagating in a cosmological background with the metric
\be\label{met} 
ds^2=-dt^2+a(t)^2dx^idx^i.
\ee
Suppose that one is interested in calculating the correlation functions of $\f$ in a suitable vacuum of the theory using the in-in formalism. Since the action is quadratic in the $\chi$ field, one may temp to integrate it out completely in the in-in path integral. Schematically written, the relevant part in the path integral involving the $\chi$ field takes the form (see e.g. \cite{w2})
\be\label{p1}
\int  {\cal D}\chi_*\,D\chi^+\, D\chi^-\,DP_\chi^+\,DP_\chi^- \exp\left\{i\left[\int_{t_0}^{t_*} \left(P_\chi^+\dot{\chi}^+-H^+\right)-(+\leftrightarrow -)\right]\right\}\Psi[\chi^+(t_0)]\overline{\Psi}[\chi^-(t_0)],
\ee
where $H$ is the Hamiltonian and  $\Psi$ denotes the vacuum wave-functional\footnote{In an interacting theory, the vacuum wave-functional may not be written as the product of $\f$ and $\chi$ pieces. However, this will be the case for us since  we eventually examine the $t_0\to-\infty$ limit.} of the $\chi$ field defined at an initial time $t_0$. Since it is very difficult to determine the vacuum of the interacting theory, one usually takes $t_0\to-\infty$ (when it is possible to do so), and in that case the sole effect of the vacuum wave-functionals in \eq{p1} is to produce the necessary $i\e$ terms for the propagators in the perturbation theory \cite{w1,w2}. The standard in-in path integral measure amounts to summing over all doubled phase space fields $\chi^\pm$ and $P_\chi^\pm$, where the fields  $\chi^\pm$ are constrained to satisfy $\chi^+(t_*)=\chi^-(t_*)=\chi(t_*)$ and  ${\cal D}\chi_*$ denote the integration over these $\chi(t_*)$ configurations defined at the fixed time $t_*$. 

One can carry out the quadratic integrals over momenta $P_\chi^\pm$ in a straightforward way. Defining a new field as 
\be \label{mu}
\m=a^{3/2}\chi,
\ee
one can see that the scale factor $a(t)$ dependent factor coming to the measure from the Gaussian $P_\chi^\pm$ integrations is exactly canceled out by the Jacobian of  the transformation \eq{mu}. As a result, the integral \eq{p1} becomes 
\be\label{p2}
\int  {\cal D}\m_*\,D\m^+\,D\m^-\,\exp\left[i(S^+-S^-)\right],
\ee
where 
\be\label{ma}
S=\fr12\int d^4x\,\left[\dot{\m}^2-\fr{1}{a^2}(\del\m)^2+\m^2\left(\fr94H^2+\fr32\dot{H}-\tilde{m}^2-g^2\f^2\right)\right].
\ee
In \eq{p2}, the vacuum wave-functionals are suppressed since their only role is to prescribe the propagators of the theory. Note that, here $\f$ plays the role of an external field, which will be  integrated eventually to give the $\f$ correlation functions. 

Defining 
\be
{\bf \upmu}=\left[\begin{array}{c}\m^+\\ \m^-\end{array}\right]
\ee
and 
\be\label{op1}
{\bf L}_s =\left[\begin{array}{cc}L&0\\0&-L\end{array}\right]+s\left[\begin{array}{cc}g^2(\f^+)^2&0\\0&-g^2(\f^-)^2\end{array}\right],
\ee
wehere
\be\label{l}
L=\fr{\del^2}{\del t^2}-\fr{1}{a^2}\del_i^2-\fr94 H^2-\fr32\dot{H}+\tilde{m}^2,
\ee
\eq{p2} can be rewritten as 
\be\label{faz}
\int  {\cal D}\m_*\,D\m^+\,D\m^-\,\exp\left\{-\fr{i}{2}\int d^4x\, \upmu^T{\bf L}_1\upmu\right\}.
\ee
As discussed in \cite{a1}, the in-in path integral is actually over the fields satisfying $\mu^+(t_*)=\mu^-(t_*)$ and $\dot{\mu}^+(t_*)=\dot{\mu}^-(t_*)$. These boundary conditions  ensure the absence of surface terms that otherwise arise after integrating \eq{ma} by parts to yield \eq{faz}. Moreover, they also make the operator $\ls$ essentially self-adjoint \cite{a1}. Since $\upmu$ is a real field, the Gaussian integral \eq{faz} gives $C[\det {\bf L}_1]^{-1/2}$. The normalization $C$ can be fixed from the requirement that the path integral yields unity when $\f=0$, which evaluates \eq{faz} as  
\be
\left[\fr{\det {\bf L}_1}{\det {\bf L}_0}\right]^{-1/2}.
\ee
One can now use \eq{det2} to express this ratio of determinants to reach 
\be
\exp\left\{-\fr12 \int_0^1ds\int d^4x\,\left[G^{++}_s[\f^+,\f^-](x;x)g^2 \f^+(x)^2-G^{--}_s[\f^+,\f^-](x;x)g^2\f^-(x)^2\right]\right\},
\ee
where $G_s^{++}$ and $G_s^{--}$ are the matrix entries of the Green function ${\bf G}_s$ of the operator \eq{op1} as defined in the appendix. Note that these Green functions depend on $\f^\pm$ in a very nontrivial way. As a result, integrating out the  $\chi$ field gives the following bare {\it in-in effective action}
 \be\label{seff}
S_{eff}[\f^+,\f^-]=\fr{i}{2} \int d^4x\,\int_0^1ds\,g^2\left[G^{++}_s[\f^+,\f^-](x;x) \f^+(x)^2-G^{--}_s[\f^+,\f^-](x;x)\f^-(x)^2\right],
\ee
which must be though to correct the classical $\f$ action given in \eq{a}. 

For given $\f^\pm$ fields, one can in principle solve for the Green function ${\bf G}_s$, which is assumed to be fixed uniquely once the vacuum of the theory is specified. However, it is difficult, if not impossible, to carry out this computation for arbitrary fields. Therefore, we consider constant $\f^\pm$ configurations to read the quantum effective potential. In that case, $\m$ becomes a free massive scalar field propagating in \eq{met} and it is easy to calculate the corresponding Green functions. Indeed, quantizing the field by introducing the standard ladder operators and the mode functions as
\be
\mu_s=\fr{1}{(2\pi)^{3/2}}\int\,d^3k\,\left[e^{i\vec{k}.\vec{x}}\,\mu_k[t,s,\f]\,a_{\vec{k}}+e^{-i\vec{k}.\vec{x}}\,\mu_k[t,s,\f]^*\,a^\dagger_{\vec{k}}\right]
\ee
where $[a_{\vec{k}},a^\dagger_{\vec{k'}}]=\d^3(\vec{k}-\vec{k'})$  and 
\bea
&&\ddot{\m}_k+\fr{k^2}{a^2}\,\m_k+\left[\tilde{m}^2+s\,g^2\f^2-\fr94 H^2-\fr32\dot{H}\right]\m_k=0,\nn\\
&&\m_k\dot{\m}_k^*-\m_k^*\dot{\m}_k=i,\label{mf}
\eea
the Green functions are determined by the following vacuum expectation values of the operators 
\be
G_s^{++}=i\left<T(\m_s^+\m_s^+)\right>,\hs{5}G_s^{--}=i\left<\overline{T}(\m_s^-\m_s^-)\right>,
\ee
where $T$ and $\overline{T}$ refer to the time and anti-time ordering operations, and the vacuum is defined as usual by $a_{\vec{k}}|0\hs{-1}>=0$. Using these in \eq{seff}, one finds for constant $\f^\pm$ that 
\be\label{seff2} 
S_{eff}[\f^+,\f^-]= \int d^4x\,a^3\,\left[V_{eff}(\f^+)-V_{eff}(\f^-)\right],
\ee
where the bare effective potential is given by\footnote{The computation of the effective potential can also be done by first introducing the Fourier modes in the path integral, which factorizes \eq{p2} for each momentum mode $\vec{k}$. This yields the determinant of an ordinary differential operator for each $\vec{k}$, which can be calculated using \eq{det}. The final result involves a sum over the momentum modes that appears as the momentum integral in \eq{veff}.}
\be\label{veff}
V_{eff}^{B}(\f)=\fr{1}{2a(t)^3}\,g^2\,\f^2\,\int \fr{d^3k}{(2\pi)^3}\int_0^1ds\,\left|\m_k[t,s,\f]\right|^2.
\ee
Recall that the mode functions $\mu[t,s,\f]$ obey \eq{mf} and they are uniquely fixed when the vacuum is specified. Using \eq{mu}, one can also express $V_{eff}^B(\f)$ using the $\chi$ mode functions by $\m_k= a^{3/2}\chi_k$ so that the initial scale dependent factor in \eq{veff} drops out. As discussed in \cite{sp1,sp2}, the classical configurations arise as  stationary phases of the  in-in path integrals. Consequently, after a suitable renormalization $V_{eff}^B(\f)$ must be thought to modify the tree-level potential $v(\f)$ given in \eq{pot} governing  the classical evolution of the $\f$ field. 

The effective potential $V_{eff}^B(\f)$ explicitly depends on time $t$ and this dependence is not surprising for a physical quantity calculated in a time evolving background. The time in \eq{veff} is naturally specified by the  correlation function of interest that is aimed to be determined in the beginning. Indeed, the time integrals in \eq{seff} or \eq{seff2} are also  limited by this initially assigned time parameter. Unlike the in-out computations in flat space that extends in the whole timeline, in the in-in framework one may imagine integrating out the ``modes'' from some initially prescribed time $t_0$ to the time of interest $t$. 

\begin{figure}
\centerline{
\includegraphics[width=6cm]{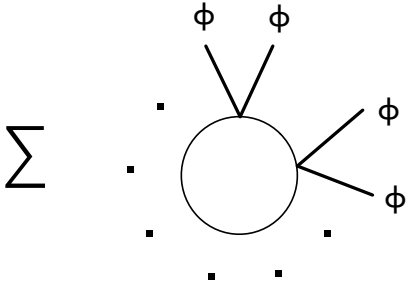}}
\caption{The perturbative one-loop graphs arising in the calculation of the effective potential \eq{veff}.} 
\label{fig1}
\end{figure}

In carrying out the $\chi$ path integral above no approximation is utilized. Consequently, given the initial interaction potential \eq{pot}, the result \eq{veff} is {\it exact}. On the other hand, one may attempt for an alternative but  perturbative path integral calculation that can be organized in powers of the external $\f$ field, where the corresponding series can be pictured as in Fig. \ref{fig1}. The effective potential  \eq{veff} equals the sum of this infinite series and conversely the infinite series can be recovered from \eq{veff} by expanding it in powers of $\f$. 

For $g\f\ll H$, the $\f$ dependence in the mode equation \eq{mf} becomes negligible. In this limit, one finds  $V_{eff}^B\simeq\fr12 g^2\f^2\left<\chi^2\right>$, which is identical to the one-loop Hartree approximation applied to the original potential \eq{pot}. 

It is instructive to use the formula \eq{veff} in {\it flat} spacetime. Using the mode function 
\be\label{fm}
\m_k=\fr{1}{\sqrt{2\o_k}}\,e^{-i\o_k t},
\ee
where $\o_k^2=k^2+\tilde{m}^2+sg^2\f^2$, in \eq{veff} and carrying out the $s$-integral, one finds
\be\label{cw1}
V_{eff}^B(\f)=\fr{1}{4\pi^2}\,g^2\f^2\,\int_0^\infty  dk\,k^2\left[\sqrt{k^2+g^2\f^2}-k\right].
\ee
In \cite{cw}, the bare Coleman-Weinberg potential for the interaction $\l\f^4/4!$ is given by 
\be\label{cw2}
V_{eff}^{CW}=\fr12\int \fr{d^4 k_E}{(2\pi)^2}\ln\left[1+\fr{\l\f_B^2}{2k_E^2}\right].
\ee
In obtaining \eq{cw2}, one defines a fluctuation field as $\f=\f_B+\d\f$, which gives a quadratic potential $\fr14 \l\f_B^2\d\f^2$, and integrates over $\d\f$. By comparing this quadratic potential with the potential we are dealing with, i.e. $\fr12g^2\f^2\chi^2$, one sees that \eq{cw2} must be identical to \eq{cw1} after the replacements $\f_B\to\f$ and $\l\to 2g^2$ (the fields that are integrated out are $\d\f$ and $\chi$, respectively). Indeed, defining $k_E=(\vec{k},k_4)$ (so that $k_E^2=\vec{k}.\vec{k}+k_4^2=k^2+k_4^2$) and carrying out the $k_4$ integral  in \eq{cw2} exactly yields \eq{cw1}. 

The form of the bare effective potential \eq{veff} is suitable for adiabatic regularization. The adiabatic mode function is defined as
\be\label{mad}
\m^{ad}_k=\fr{1}{\sqrt{2\O_k}}\exp\left[-i\int^t \O_k(t')dt'\right]
\ee
and the mode equation \eq{mf} implies 
\be\label{oe}
\O_k^2=\o_k^2+\fr34\fr{\dot{\O}_k^2}{\O_k^2}-\fr12\fr{\ddot{\O}_k}{\O_k},
\ee
where
\be\label{wk}
\o_k^2=\fr{k^2}{a^2}+\tilde{m}^2+s g^2 \f^2.
\ee
Note that \eq{mad} obeys the Wronskian condition identically. One may now solve \eq{oe} iteratively in the number of time derivatives starting from the 0th order solution $\O_k^{[0]}=\o_k$. Truncating the iterative solution at any desired order\footnote{In some cases, there can be an inherent ambiguity in defining the order in this iterative solution, see \cite{amb}.} defines an approximate adiabatic vacuum of that order.  

Since \eq{veff} is expected to be quadratically divergent (this is the case for the Bunch-Davies vacuum), it is enough to subtract the second order adiabatic terms from \eq{veff}. A straightforward calculation gives the renormalized effective action as 
\be\label{nveff}
V_{eff}(\f)=\fr{1}{2a(t)^3}\,g^2\,\f^2\,\int \fr{d^3k}{(2\pi)^3}\int_0^1ds\,\left[\left|\m_k[t,s,\f]\right|^2-\left|\m_k^{ad}{}^{(2)}\right|^2\right],
\ee
where 
\be\label{ad2}
\left|\m_k^{ad}{}^{(2)}\right|^2=\fr{1}{2\o_k}\left[1+\fr98\fr{H^2}{\o_k^2}+\fr34\fr{\dot{H}}{\o_k^2}-\fr58\fr{H^2k^4}{a^4\o_k^6}-\fr14\fr{\dot{H}k^2}{a^2\o_k^4}+\fr12\fr{H^2k^2}{a^2\o_k^4}\right].
\ee
Note that $\o_k$, which is defined in \eq{wk}, depends on the parameter $s$. Eq. \eq{nveff} is the main result of this section. 

The subtracted terms in \eq{nveff} can be thought to arise from a counterterm  potential $\d V(\f)$ appearing in the bare action. It can be found as 
\be
\d V(\f)=-a_1g^4\f^4-2a_1 g^2\mt^2\f^2-g^2\f^2\left[a_2H^2+a_3\dot{H}\right],
\ee
where 
\bea 
&&a_1=\fr{1}{16\pi^2}\int_0^\infty \fr{k^2dk}{(k^2+1)^{1/2}},\nn\\
&&a_2=\fr{9}{64\pi^2}\int_0^\infty \fr{k^2\,dk}{(k^2+1)^{3/2}}+ \fr{1}{16\pi^2}\int_0^\infty \fr{k^4\,dk}{(k^2+1)^{5/2}}-\fr{5}{64\pi^2}\int_0^\infty \fr{k^6\,dk}{(k^2+1)^{7/2}},\\
&&a_3=\fr{3}{32\pi^2}\int_0^\infty\fr{k^2\,dk}{(k^2+1)^{3/2}}-\fr{1}{32\pi^2}\int_0^\infty\fr{k^4\,dk}{(k^2+1)^{5/2}}.
\eea
This interpretation justifies the adiabatic subtraction terms in \eq{nveff} since the regularization procedure can be recast as a standard renormalization method (see \cite{y5} for a similar renormalization treatment involving adiabatic  subtractions).  

One may see that in the flat space limit \eq{nveff} gives $V_{eff}(\f)=0$. This can be viewed as a prescription yielding an unambiguous renormalized result. Namely, the  inherently present finite renormalizations are fixed in such a way that as the expansion of the universe is switched off the quantum effective potential vanishes. Thus, the adiabatic regularization physically captures the quantum effects directly related to the cosmic expansion or ``particle creation". 

\section{\leftline{Applications to inflation}}

It is straightforward to apply the above formalism to any given cosmological background and in this section we consider the standard single field slow-roll inflationary scenario. Let $\f$ be the inflaton field and $\chi$ be another scalar field coupled to $\f$ in the form \eq{pot}. Such couplings are necessary for the decay of the inflaton during reheating and the potential \eq{pot} is a plausible alternative, see e.g. \cite{reh1,reh2}. Although, at the classical level the $\chi$ field does not affect the background dynamics, one may wonder how the quantum $\chi$ fluctuations modify the classical inflaton potential $v(\f)$ through the coupling \eq{pot}. For large field models like chaotic inflation, where the value of the inflaton field background exceeds the Planck scale, the quantum corrections might be important. 

In this computation, the slow-rolling of the inflaton field gives sub-leading corrections and therefore we set $a=\exp(Ht)$. The Bunch-Davies mode function of the $\chi$ field in de Sitter space obeying \eq{mf} is given by 
\be
\m_k[t,s,\f]=e^{i\pi\n/2}\sqrt{\fr{\pi}{4H}}H^{(1)}_\n\left(ke^{-Ht}/H\right),
\ee
where $H^{(1)}_\nu$ is the standard Hankel function of first kind and 
\be
\n=\sqrt{\fr94-\fr{sg^2\f^2+\mt^2}{H^2}}.
\ee
Using the mode function in \eq{nveff} and determining the adiabatic subtraction terms for de Sitter space from \eq{ad2}, we obtain 
\bea
V_{eff}(\f)=\fr{H^4}{4\pi^2}&&\left(\fr{g^2\f^2}{H^2}\right)
\int_0^\infty u^2\,du\int_0^1 ds\nn\\
&&\left[e^{-\pi\textrm{Im}(\nu)}\fr{\pi}{4}\left|H^{(1)}_\n(u)\right|^2-\fr{1}{2\o(u)}\left(1+\fr{9}{8\o(u)^2}+\fr{u^2}{2\o(u)^4}-\fr{5u^4}{8\o(u)^6}\right)\right],\label{son}
\eea
where 
\be
\o(u)^2=u^2+\fr{sg^2\f^2+\mt^2}{H^2}.
\ee
Eq. \eq{son} gives the fully renormalized exact quantum effective potential for the inflaton field that arises from its interaction with the $\chi$ field with the potential \eq{pot}. In de Sitter space, $V_{eff}(\f)$ turns out to be time independent due to the underlying symmetries of the background.  

If $\mt\not=0$, the $u$-integral is convergent near $u=0$ for any value of $\f$. When $\mt=0$, i.e when $\chi$ is massless, one may see that as $\f\to0$ the integrand in \eq{son} vanishes and thus $V_{eff}(\f)\to0$. On the other hand, for nonzero $\f$ the $u$-integral is again convergent as $u\to0$ since $\textrm{Re}(\nu)<3/2$. Moreover, the adiabatic regularization guarantees UV finiteness and thus the convergence as $u\to\infty$. Therefore, \eq{son} is completely safe both at IR and UV, corresponding to $u\to0$ and $u\to\infty$ limits, respectively. 

As mentioned above, \eq{son} is the quantum correction to the classical inflaton potential due to its coupling to the $\chi$ field and it can be used in Einstein's equations to modify the classical evolution (although the corrections turn out to be small as we will show below). This effect is somehow complementary to the one obtained by the so called tadpole method, where one defines a classical background and a fluctuation field, and the evolution of the background is fixed by the vanishing of the tadpole of the fluctuation field, see e.g. \cite{tad1,tad2,tad3}. Indeed, one may see that at one loop the tadpole method applied to the interaction potential in \eq{pot} is equivalent to the Hartree approximation, where $g^2<\hs{-1}\chi^2\hs{-1}>$ appears as an effective mass term for the inflaton.  At higher loops, the tadpole method necessarily includes the inflaton loops which are completely absent in \eq{son}. 

In a single field model, it is natural to assume that $\mt\ll H$, since otherwise the $\chi$ field is expected to alter the inflationary background evolution unless the initial conditions are finely tuned. Neglecting $\mt$ in \eq{son}, one sees that the effective potential becomes $H^4$ times a dimensionless function $f(g^2\f^2/H^2)$. For $g\f\sim H$, the function $f$ takes values of order unity and its dependence on its argument is somehow weak.  In Table \ref{tb1}, we numerically evaluate $V_{eff}$ for various values of $g\f/H$. Since $V_{eff} \propto H^4$ and the background energy density is of the order of $H^2M_p^2$, the corrections induced by $V_{eff}$ on the cosmological observables are suppressed by the factor $H^2/M_p^2$.  
 
\begin{table}
\begin{center}
\begin{tabular}{| c | |c | c|c|c|c|c|c|c|}
    \hline
    $g^2\f^2$&0.01&0.1&1 &2&4&20&100&200  \\ \hline
     $V_{eff}$ & 0.005&0.03&0.06&  0.07&0.07 &0.08 &0.07&0.06\\
    \hline
  \end{tabular}
\end{center}
\caption{The numerical values of \eq{son} for various field values in units of the Hubble parameter $H$.}\label{tb1}
\end{table}

Let us now discuss a similar problem involving the cosmological perturbations. We focus on the curvature perturbation $\z$, which can be introduced in the metric as 
\be\label{mz}
ds^2=-dt^2+a(t)^2\,e^{2\z(t,\vec{x})}dx^idx^i.
\ee
As before, we treat $\z$ as an {\it external field} and integrate out  the $\chi$ field  in the path integral to obtain an effective action. In the presence of the curvature perturbation (in the $\delta\f=0$ gauge) the metric \eq{mz} must include nontrivial lapse $N$ and the shift $N^i$ functions that depend on $\z$ \cite{mal}. Nevertheless, we take $\z$ to be an (off-shell) superhorizon perturbation and apply a derivative expansion to the effective $\z$ action to determine an effective potential. In that case $\z$ dependent factors in both $N$ and $N^i$ become negligible at the leading order since they always  contain derivatives. We take $\chi$ to be a minimally coupled massless field propagating in the metric \eq{mz} where the action becomes
\be\label{acz}
S=\fr12 \int d^4x\,a^3\,e^{3\z}\left[\dot{\chi}^2-\fr{1}{a^2}e^{-2\z}(\del\chi)^2\right].
\ee
Note that there is a {\it shift symmetry} in the action that is given by 
\bea\label{sy}
&&\z\to\z+\l,\\
&&x^i\to e^{-\l}x^i,\nn
\eea
and we demand this symmetry to be preserved in our computation below.   

The path integral over  the momentum variable $P_\chi$  is nontrivial since it involves the external $\z$ field. After applying the following canonical transformation 
\bea
&&\m=a^{3/2}\exp\left[3\z/2\right]\chi,\nn\\
&&P_\m=a^{-3/2}\exp\left[-3\z/2\right]P_\chi,
\eea
which preserves the measure 
\be
D\chi\, DP_\chi=D\m \,DP_\m,
\ee
the path integral over $P_\m$ becomes independent of $\z$ and decouples. As a result, one is left with an integral over $\m$ with the action
\bea\label{acm}
S=\fr12\int\,d^4x\,\left[\dot{\m}^2+\left(\fr32\dot{H}+
\fr94 H^2\right)\m^2-\fr{e^{-2\z}}{a^2}(\del\m)^2\right],
\eea
where the derivatives of $\z$ are again neglected for consistency with the original action \eq{acz}. This final path integral yields the functional determinant of the operator in \eq{acm}. It can be seen that the in-in calculation is again separated into two identical $+$ and $-$ branches, as in the previous section. 

Even for a slowly changing external field $\z$, it is very difficult to calculate this functional determinant exactly. On the other hand, 
the constant $\z$ configurations are pure gauge due to the shift symmetry and the exactly calculable determinant does not give any information about the effective action. Therefore, we focus on the curvature perturbations obeying  
\bea
&&\dot{\z}\ll H,\nn\\
&&\del_i\z\ll  k_*\z\label{cz1}
\eea
for some $k_*$ so that the path integral can partially be carried out  as follows: One may see that the time derivatives of $\z$ always appear with $H$ in the combination $(\dot{\z}+H)$, thus the first condition in \eq{cz1} allows one to neglect $\dot{\z}$ in the action. Decomposing the path integral measure in the momentum space as 
\be
D\mu=\prod_k d\m_k=\left(\prod_{k<k_*} d\m_k\right)\left(\prod_{k>k_*}d\m_k.
\right)
\ee
and integrating out the modes with $k>k_*$, the spatial derivatives of $\z$ can also be neglected in the action. As a result, $\z$ can effectively be treated as a constant in such a partial path integral of ``high energy modes".  For an actual superhorizon curvature perturbation $\zs$, both conditions in \eq{cz1} are satisfied once  $k_*$ is taken to be the comoving horizon scale determined by the metric \eq{mz}. Therefore, by setting 
\be
k_*=e^{\zs}a(t)H
\ee
it is possible to obtain an effective potential for $\z_{sh}$ by integrating out the subhorizon $\chi$ modes. 

In normalizing this path integral, one must preserve the original shift symmetry \eq{sy} (this is necessary due to the way $\z$ is introduced in the metric \eq{mz}). For that it is convenient to calculate the derivative of the functional determinant from \eq{deski} (with the $s$ derivative replaced by a $\zs$ derivative), which yields the derivative of the effective potential. The undetermined additive constant of integration can uniquely be fixed by demanding the symmetry \eq{sy}. A straightforward calculation using the formulas derived in the previous section gives an effective bare potential for $\zs$ as 
\be\label{pz}
V_{eff}^{B}(\zs)'=-\fr{1}{a^5}e^{-2\zs}\int_{k>k_*}\fr{d^3k}{(2\pi)^3}\,\,k^2\,\left|\m_k\right|^2,
\ee
where the prime denotes derivative with respect to argument and the mode functions obey
\bea
&&\ddot{\m}_k+e^{-2\zs}\fr{k^2}{a^2}\,\m_k-\left[\fr94 H^2+\fr32\dot{H}\right]\m_k=0,\nn\\
&&\m_k\dot{\m}_k^*-\m_k^*\dot{\m}_k=i.\label{mf2}
\eea
We note that \eq{pz} is exact to the leading order in the derivative expansion. 

In the Bunch-Davies vacuum, the bare potential \eq{pz} has a quartic UV divergence that can be cured by adiabatic regularization  by subtracting the terms up to fourth order. This gives 
\be\label{pz2}
V_{eff}(\zs)'=-\fr{1}{a^5}e^{-2\zs}\int_{k>k_*}\fr{d^3k}{(2\pi)^3}\,\,k^2\,\left[\left|\m_k\right|^2-\left|\m_{ad}^{(4)}\right|^2\right],
\ee
where
\be
\left|\m_{ad}^{(4)}\right|^2=e^{\zs}\fr{a}{2k}\left[1+\fr{e^{2\zs}}{2k^2}\left(\dot{a}^2+a\dot{a}\right)+\fr{e^{4\zs}}{8k^4}\left(3\dot{a}^4+3a\dot{a}\ddot{a}-5a^2\dot{a}a^{(3)}-a^3a^{(4)}\right)\right].
\ee
One may see that in the exact de Sitter space \eq{pz2} vanishes,\footnote{In applying the adiabatic regularization to massless fields, one must actually start with a massive field and then take a limit. In this procedure, one discovers some terms that survive ``mass going to zero" limit when the momentum integral is unconstrained and this is how the standard trace anomaly is recovered in the adiabatic regularization \cite{trace}. In our case, since the momentum integral in \eq{pz2} has already an IR cutoff, this procedure gives no extra term in the massless limit.}  which is a standard result in adiabatic regularization (see e.g. \cite{park2}). Therefore,  the slow rolling of the inflaton must be taken into account and here we take the background to be the space with constant deceleration obeying
\be
\dot{H}=-\e\,H^2,
\ee
where $\e$ is a constant that is not necessarily small. Introducing the conformal time $ad\eta=dt$, the scale factor can be expressed as 
\be
a=(-H_0\eta)^{-1/(1-\e)}
\ee
where the constant $H_0$ can be identified with the inflationary Hubble scale when $\e\ll1$. The Bunch-Davies mode function obeying \eq{mf2} is given by
\be
\mu_k=\sqrt{\fr{\pi}{4H_0}}a^{\e/2}H_\nu^{(1)}\left(-e^{-\zs}k\eta\right),\hs{10}\nu=\fr{3-\e}{2(1-\e)} 
\ee
and \eq{pz2} becomes
\be\label{ps}
V_{eff}(\zs)=C\,H_0^4\,e^{3\zs}
\ee
where 
\be\label{cc}
C=-\fr{1}{6\pi^2}\int_1^\infty\,du\,u^4\left[\fr{\pi}{4}a^\e|H_\nu^{(1)}(ua^\e)|^2-\fr{1}{2u}\left(1+\fr{2-\e}{2(1-\e)^2u^2a^{2\e}}+\fr{3\e(2-\e)(3-2\e)}{8(1-\e)^4u^4a^{4\e}}\right)\right].
\ee
In fixing the integration constant in \eq{ps} we demand the action to have the symmetry \eq{sy}. Note that the coefficient $C$ is time dependent, but this dependence is mild in the slow-roll regime.  

\begin{table}
\begin{center}
\begin{tabular}{| c | |c | c|c|c|c|c|c|c|c|}
    \hline
    $\e$&$10^{-1}$&$10^{-1}$&$10^{-1}$ &$10^{-2}$&$10^{-2}$&$10^{-2}$&$10^{-3}$&$10^{-3}$  &$10^{-3}$ \\ 
    \hline
     $a$ & 1&10 &100 & 1&10 &100 &1 &10&100\\
    \hline
    C  & $10^{-3}$&$4\times 10^{-4}$&$ 10^{-4}$& $1.1\times 10^{-4}$&$1.0\times 10^{-4}$ &
    $0.9\times 10^{-4}$ &$1.10\times 10^{-5}$&$1.08\times 10^{-5}$&$1.07\times 10^{-5}$\\
    \hline 
  \end{tabular}
\end{center}
\caption{The numerical values of the coefficient $C$ given in \eq{cc} for different values of $\e$ evaluated at different times labeled by the scale factor $a$.  Note that for this set $C\sim 10^{-2}\e$ and the pre-factor is related to the overall number $1/(6\pi^2)$ appearing in \eq{cc}.}\label{tb2}
\end{table}

One may see that $C\to0$ as $\e\to0$. Thus, $C\simeq \e$ for $\e\ll1$ (see Table \ref{tb2}) and in the slow-roll approximation \eq{pz2} becomes
\be\label{pe}
 V_{eff}(\zs)\simeq \e\,H_0^4\,e^{3\zs}.
\ee
It is important to emphasize that \eq{pe} can only be used for superhorizon modes and the corrections induced by \eq{pe} occur after horizon crossing. Physically, one may imagine them to arise due to the scatterings of the superhorizon $\zs$ with the subhorizon (virtual) $\chi$ fluctuations. The theorems proved in \cite{ c1,c2} regarding the conservation of superhorizon $\z$ in quantum theory are not applicable here due to presence of the extra scalar field $\chi$. 

One may calculate the tree level corrections to the power spectrum and non-gaussianity that follow from \eq{pe}. The quadratic piece coming from the expansion of the exponential yields a mass term that modifies the power spectrum and the cubic term yields a nonzero $f_{NL}$. The relative magnitude of the  correction to the power spectrum and $f_{NL}$ are given by 
\be
N\,\fr{H^2}{M_p^2}
\ee
where $N$ is the number of e-folds from the horizon crossing (of the mode of interest) till the end of inflation (with the start of reheating loop corrections may still be important as discussed in \cite{a2,a3,a4} but the whole calculation takes a different form). This slight enhancement is similar to  $\ln(a)$ factors appearing in loop corrections, see e.g. \cite{ir1}.  
 
Finally, it is also possible to work out an effective potential for the inflaton that arises from integrating out its own quantum fluctuations. To see that, one may introduce the standard generating functional\footnote{It is straightforward to repeat the analysis below for the in-in case.}
\be\label{gf}
e^{iW[J]}=\int D\f \,e^{iS_{cl}[\f]+i\int J\f},
\ee
and the quantum effective action
\be
\Gamma[\f_{cl}]=W[J]-\int J\f_{cl}.
\ee
Using $\d\Gamma/\d \f_{cl}=-J$, \eq{gf} can be written as 
\be
e^{i\Gamma[\f_{cl}]}=\int D\f\, e^{iS_{cl}[\f]-i\int \fr{\d\Gamma}{\d\f_{cl}}(\f-\f_{cl})},
\ee
which is an exact expression for $\Gamma$. Expanding the quantum effective action as 
\be
\Gamma=S_{cl}+S^{(1)},
\ee
where $S^{(1)}$ denotes a first order (one-loop) correction,  defining a new integration variable $\chi=\f-\f_{cl}$  and keeping only the leading order terms one finds 
\be
e^{iS^{(1)}[\f_{cl}]}\simeq\int D\chi\, \exp\left[\fr{i}{2}\int \chi \fr{\d^2S_{cl}}{\d\f_{cl}\d\f_{cl}}\chi\right].
\ee
This gives the one-loop effective action $S^{(1)}$ in terms of the functional determinant of the (second order) fluctuation operator. Starting with a classical inflaton potential $v(\f)$, the adiabatically regularized one loop  quantum potential is given by \eq{nveff} with the replacement  
\be
g^2\to v''(\f)
\ee
where the prime denotes derivative with respect to argument. A similar analysis can be done for the curvature perturbation $\z$. 

\section{\leftline{Conclusions}}

In this paper, we use mode functions to express the functional determinants that give one-loop effective potentials of scalar fields propagating in cosmological backgrounds and apply adiabatic regularization to cure the resulting infinities. It turns out that in this problem the adiabatic regularization can be cast into a well defined renormalization procedure corresponding to the specific counterterms added to the action. The finite renormalizations are automatically fixed in such a way that the effective potential vanishes when the expansion of the universe is turned off, thus it captures the quantum effects related to the expansion. 

The same method can also be applied to derive an effective potential for the superhorizon curvature perturbation $\zs$ encoding its interactions with virtual subhorizon modes circulating in the loops. For a given mode, the effective potential can be used starting from the horizon crossing  time till the beginning  of reheating. In that case, it is difficult to interpret the adiabatic regularization by counterterms added to the action since the momentum integral is cut off at the horizon crossing time (see e.g. slow-roll parameter dependent terms in \eq{cc}).  In any case, the loop effects involving cosmological perturbations already contain non-renormalizable  gravitational interactions and consequently  the whole problem should be viewed in an effective field theory approach. 

The corrections induced by the effective inflaton potential on the background evolution and effective $\zs$ potential on the cosmological observables like power spectrum and non-gaussianity turn out to be small even in the large field models like chaotic. These findings support the general view that inflation can fully be understood in a semiclassical approach and quantum backreaction effects can safely be neglected.

\appendix*

\section{The functional determinant of a time dependent  in-in operator}\label{ap1}

We consider the Hilbert space of {\it doublets} of time dependent functions 
\be
\Phi(t)=\left[\begin{array}{c}\phi^+(t)\\\phi^-(t)\end{array}\right],
\ee
which are defined in the interval $(t_a,t_b)$ with the inner product
\be\label{rip}
<\hs{-1}\Phi_1|\Phi_2\hs{-1}>=\int_{t_a}^{t_b} dt \,\Phi_1^T(t) \Phi_2(t)=\int_{t_a}^{t_b} dt \,\left[\phi_1^+(t) \phi_2^+(t)+\phi_1^-(t) \phi_2^-(t)\right].
\ee
We assume that these functions obey suitable boundary conditions so that the one parameter family of operators ${\bf L}_s$, which have the following form  
\be\label{jj}
{\bf L}_s =\left[\begin{array}{cc}L&0\\0&-L\end{array}\right]+s\left[\begin{array}{cc}\O^+(t)&0\\0&-\O^-(t)\end{array}\right],
\ee
are self-adjoint with {\it discrete spectrum}. Here, $L$ denotes a generic second order (ordinary) differential operator, $s$ is a real parameter and $\O^{\pm}$ are smooth functions. Our aim is to calculate the ratios of the two determinants in this family, i.e. $\det {\bf L}_1/\det {\bf L}_0$. 

For a given parameter $s$, let 
\be\label{ecom}
y_n(s)=\left[\begin{array}{c}y_n^+(t,s)\\y_n^-(t,s)\end{array}\right]
\ee
to denote the eigenfunctions of the self-adjoint operator \eq{jj}, i.e. 
\be
\ls y_n(s)=\l_n(s)y_n(s). 
\ee
The eigenfunctions can be taken to be orthonormal 
\be
<\hs{-1}y_n(s)|y_m(s)\hs{-1}>=\d_{nm}
\ee
and they also form a complete set of basis functions in the Hilbert space satisfying the following completeness relation 
\be
\sum_n y_n(s,t)y_n^T(s,t') =\left[\begin{array}{cc}\d(t-t')&0\\0&\d(t-t')\end{array}\right].
\ee
The Green function with the prescribed boundary conditions is defined by  
\be
{\bf L}_s{\bf G}_s=\left[\begin{array}{cc}\d(t-t')&0\\0&\d(t-t')\end{array}\right].
\ee
We assume that $\ls$ has {\it no zero modes} and the Green function is uniquely defined. The matrix entries, which can be introduced as 
\be\label{g1}
{\bf G}_s(t,t')=\left[\begin{array}{cc}G^{++}_s(t,t')&G^{+-}_s(t,t')\\ G^{-+}_s(t,t')&G^{--}_s(t,t')\end{array}\right],
\ee
are fixed by the eigenfunctions $y_n(s)$ as  
\be\label{g2}
{\bf G}_s(t,t')=\sum_n \fr{y_n(s,t)y_n^T(s,t')}{\l_n(s)}.
\ee
As discussed in \cite{a1}, one may use this identity in constructing standard in-in propagators in flat space. 

Formally, the determinant can be written as
\be 
\det {\bf L}_s= \exp\left[{\textrm Tr} \ln{\bf L}_s\right],
\ee
where the trace is defined in the Hilbert space \eq{rip}. Differentiating the above expression with respect to $s$ and noting that ${\bf L}_s^{-1}={\bf G}_s$  one  finds 
\be\label{deski}
\fr{d}{ds}\ln\det{\bf L}_s={\textrm Tr}\left[{\bf G}_s\fr{d{\bf L}_s}{ds}\right],
\ee
which can be used to express the determinant an operator in terms of its Green function (see e.g. \cite{pi1,pi2,pi3,pi4,pi5,pi6}). 

To get a more rigorous derivation of \eq{deski}, one may follow a simple method presented in \cite{kc}. The determinant of the operator, which is defined as
\be
\det\ls=\prod_n\l_n(s),
\ee
generically diverges. However, the ratio of the two determinants like $\det {\bf L}_1/\det {\bf L}_0$ is usually well defined and finite. Differentiating the eigenvalue equation
\be
\l_n(s)=<\hs{-1}y_n(s)|\ls|y_n(s)\hs{-1}>
\ee
with respect to $s$ (an $s$-derivative is denoted by a prime), one may find 
\be
\l_n(s)'=<\hs{-1}y_n(s)|\left[\begin{array}{cc}\O^+&0\\0&-\O^-\end{array}\right]   |y_n(s)\hs{-1}>.
\ee
Then,
\be
\sum_n\ln\left[\l_n(s)\right]'=\sum_n\int_{t_a}^{t_b}dt\left[\fr{y_n^+(t,s)y_n^+(t,s)}{\l_n(s)}\O^+(t)-\fr{y_n^-(t,s)y_n^-(t,s)}{\l_n(s)}\O^-(t)\right],
\ee
where the components of the eigenfunctions are introduced in \eq{ecom}. Integrating this last equation with respect ot $s$ and using  \eq{g1} and \eq{g2}, one finally obtains
\be\label{det} 
\ln\left[\fr{\det {\bf L}_1}{\det {\bf L}_0}\right]=\int_0^1ds\int_{t_a}^{t_b}dt\left[G^{++}_s(t,t)\O^+(t)-G^{--}_s(t,t)\O^-(t)\right],
\ee
which expresses the ratios of the determinants in terms of the Green function. 

One nice feature of the construction of \cite{kc} is that it can be readily generalized to partial differential operators, unlike the case with the Gel'fand-Yaglom theorem (see e.g. \cite{d1,d2}).  Consider the doublets of functions
\be
\Phi(x)=\left[\begin{array}{c}\phi^+(x^\m)\\\phi^-(x^\m)\end{array}\right],
\ee
which are  defined in an $n$-dimensional space parametrized with the coordinates $x^\m$ and endowed with  the following inner product 
\be\label{rip2}
<\hs{-1}\Phi_1|\Phi_2\hs{-1}>=\int  d^nx  \,\Phi_1^T(x) \Phi_2(x).
\ee
Consider a self-adjoint operator of the form \eq{jj} where $L$ now denotes a {\it partial differential operator} and $\O^\pm(t)\to\O^\pm(x^\m)$. It is easy to repeat the above steps to show that
\be\label{det2} 
\ln\left[\fr{\det {\bf L}_1}{\det {\bf L}_0}\right]=\int_0^1ds\int d^nx\,\left[G^{++}_s(x;x)\O^+(x)-G^{--}_s(x;x)\O^-(x)\right],
\ee
where the Green function is defined as  
\be
{\bf L}_s{\bf G}_s(x;x')=\left[\begin{array}{cc}\d^n(x-x')&0\\0&\d^n(x-x')\end{array}\right].
\ee
In the cosmological setting of our interest, one has  $x^\m=(t,x^i)$ and $L$ is fixed by the covariant Laplacian as in \eq{l}. In that case, the spectrum of the operator $\ls$ is usually  continuous. However, one can initially take the spatial coordinates $x^i$ to be periodic with period $a$ giving a discrete set of momenta  $2\pi n^i/a$ with integers $n^i$. One may then take $a\to\infty$ limit that would replace the discrete Fourier modes with the continuous Fourier transform, which would justify the use of  \eq{det2}. 

\begin{acknowledgments} 
ESK is supported by T\"{U}B\.{I}TAK-B\.{I}DEB 2211-A Fellowship. 
 \end{acknowledgments}

\end{document}